\documentclass[prd,nofootinbib,twocolumn,superscriptaddress,showpacs]{revtex4}
\usepackage{graphicx}
\usepackage{bm}
\usepackage{natbib}
\usepackage{amsfonts} 
\usepackage{amssymb}
\usepackage{bbm}
\usepackage{epsfig}
 \usepackage{epstopdf}
 \usepackage{amsmath,amssymb}
\usepackage{graphics}
\usepackage{subfigure}
\usepackage{colordvi}
\usepackage{color}

\begin{document}

\newcommand{\be}{\begin{equation}}
\newcommand{\ee}{\end{equation}}
\newcommand{\bea}{\begin{eqnarray}}
\newcommand{\eea}{\end{eqnarray}}

\newcommand{\kp}{\kappa}
\newcommand{\Om}{\Omega}
\newcommand{\de}{\Delta}
\newcommand{\eps}{\epsilon}
\newcommand{\g}{\gamma}

\newcommand{\tot}{\mathrm{tot}}
\newcommand{\ini}{\mathrm{in}}
\newcommand{\eff}{\mathrm{eff}}
\newcommand{\nr}{\mathrm{NR}}
\newcommand{\e}{\mathrm{end}}
\newcommand{\h}{\mathcal{H}}
\newcommand{\Hyd}{\mathrm{H}}
\newcommand{\He}{\mathrm{He}}
\newcommand{\eV}{\mathrm{eV}}
\newcommand{\keV}{\mathrm{keV}}
\newcommand{\GeV}{\mathrm{GeV}}
\newcommand{\lya}{Ly$\alpha \ $}
\newcommand{\iMpc}{\mbox{ Mpc$^{-1}$}}
\newcommand{\cms}{cm$^3$s$^{-1}$}
\newcommand{\sv}{\langle \sigma_{\mathrm{A}} v \rangle}
\newcommand{\ud}{\mbox{d}}
\newcommand{\fesc}{f_{\mathrm{esc}}}
\newcommand{\fa}{f_{\alpha}}
\newcommand{\fx}{f_{X}}
\newcommand{\Nion}{N_{\mathrm{i}}}
\newcommand{\sigv}{\langle \sigma_A v \rangle }
\newcommand{\ic}{\mathrm{IC}}
\newcommand{\prompt}{\mathrm{prompt}}
\newcommand{\Mmin}{M_{\mathrm{min}}}
\newcommand{\nA}{n_{\mathrm{A}}}
\newcommand{\nH}{n_{\mathrm{H}}}
\newcommand{\nHe}{n_{\mathrm{He}}}
\newcommand{\Th}{\theta_{13}}
\newcommand{\nua}{\nu_{\alpha}}
\newcommand{\fin}{\mathrm{fin}}

\newcommand{\Trh}{T_\mathrm{RH}}
\newcommand{\Neff}{N_\mathrm{eff}}
\newcommand{\MeV}{\mathrm{MeV}}

\newcommand\pp{\,\,\,.}
\newcommand\vv{\,\,\,,}

\newcommand{\apjl}{Astrophys. J. Lett.}
\newcommand{\apjs}{Astrophys. J. Suppl. Ser.}
\newcommand{\aap}{Astron. \& Astrophys.}
\newcommand{\aj}{Astron. J.}
\newcommand{\araa}{Ann. Rev. Astron. Astrophys. } 
\newcommand{\mnras}{Mon. Not. R. Astron. Soc.}
\newcommand{\physrep}{Phys. Rept.}
\newcommand{\jcap}{JCAP}

\begin{minipage}[t]{6.9in}
\hfill{\tt IFIC/15-70}
\end{minipage}

\title{Bounds on very low reheating scenarios after Planck}

\author{P.\ F.\ de Salas}
\affiliation{Instituto de F\'{\i}sica Corpuscular  (CSIC-Universitat de Val\`{e}ncia),
c/ Catedr\'atico Jos\'e Beltr\'an, 2, 46980 Paterna (Valencia), Spain}

\author{M.\ Lattanzi}
\affiliation{Dipartimento di Fisica e Scienze della Terra, Universit\`a di Ferrara and INFN, Sezione di Ferrara,
Polo Scientifico e Tecnologico - Edificio C Via Saragat, 1, I-44122 Ferrara, Italy}

\author{G.\ Mangano}
\affiliation{INFN, Sezione di Napoli, Complesso Univ. Monte S. Angelo, I-80126 Napoli, Italy}

\author{G.\ Miele}
\affiliation{INFN, Sezione di Napoli, Complesso Univ. Monte S. Angelo, I-80126 Napoli, Italy}
\affiliation{Dipartimento di Fisica {\it Ettore Pancini}, Universit\`a di Napoli Federico II, Complesso Univ. Monte S. Angelo, I-80126 Napoli, Italy}

\author{S.\ Pastor}
\affiliation{Instituto de F\'{\i}sica Corpuscular  (CSIC-Universitat de Val\`{e}ncia),
c/ Catedr\'atico Jos\'e Beltr\'an, 2, 46980 Paterna (Valencia), Spain}

\author{O.\ Pisanti}
\affiliation{INFN, Sezione di Napoli, Complesso Univ. Monte S. Angelo, I-80126 Napoli, Italy}
\affiliation{Dipartimento di Fisica {\it Ettore Pancini}, Universit\`a di Napoli Federico II, Complesso Univ. Monte S. Angelo, I-80126 Napoli, Italy}

\date{\today}

\begin{abstract}
We consider the case of very low reheating scenarios ($T_{\rm RH}\sim\mathcal{O}({\rm MeV})$) with a better calculation of the production of the relic neutrino background (with three-flavor oscillations). At 95\% confidence level, a lower bound on the reheating temperature $T_{\rm RH}>4.1$ MeV is obtained from Big Bang Nucleosynthesis, while $T_{\rm RH}>4.3$ MeV from Planck data for very light ($\sum m_i = 0.06\,\eV$) neutrinos. If neutrino masses are allowed to vary, Planck data yield $T_{\rm RH}>4.7$ MeV, the most stringent bound on the reheating temperature to date. Neutrino masses as large as 1 eV are possible for very low reheating temperatures. 

\end{abstract}

%

%

\maketitle


\section{Introduction} \label{sec:intro}

A common assumption about the history of the Universe is that its expansion was fixed by relativistic particles at early epochs. 
This radiation-dominated era usually arises as a result of the thermalization of the decay products of a massive particle, 
a process called reheating. The best example is the reheating process after primordial inflation that occurred at very large
temperatures. However, it is still possible that unstable nonrelativistic particles, other than the inflaton, were responsible of 
more than one reheating processes at different times in the evolution of the 
Universe, leading to a series of matter and radiation-dominated phases.

What one can say is that there was a final period dominated by relativistic particles in thermal equilibrium
starting at a maximum temperature $T_{\rm RH}$, at least before Big Bang Nucleosynthesis (BBN) so that
the primordial production of light elements is in agreement with the observed abundances. If one considers
$T_{\rm RH}$ as an unknown quantity that characterizes very low reheating scenarios, a lower bound on its 
possible value can be obtained from BBN. This was the subject of references 
\cite{Kawasaki:1999na,Kawasaki:2000en,Giudice:2000ex,Giudice:2000dp,Hannestad:2004px}, where 
late-time entropy production was found to be limited to $T_{\rm RH}\gtrsim 0.5-0.7$ MeV.

The main consequence of very low reheating scenarios concerns the production of neutrinos, because they
are the relativistic particles with the largest decoupling temperature. Weak processes
involving neutrinos are only effective at cosmological temperatures above 1 MeV. Therefore, for $T_{\rm RH}\sim{\cal O}({\rm MeV})$,
the thermalization of the neutrino background could be incomplete due to the lack of interactions. In such a case,
neutrino spectra would not present an equilibrium form with the same temperature as the electromagnetic plasma,
and in particular the contribution of neutrinos to the energy density of radiation, measured in terms of the parameter
$N_{\rm eff}$, would be smaller than the standard value of $3.046$ \cite{Mangano:2005cc,new_stand_Neff}. This would affect the
expansion rate during BBN, as well as the influence of electron neutrinos and antineutrinos on weak processes
relating neutrons and protons, both effects leading to the lower bound on $T_{\rm RH}$ mentioned above. The effect of flavor neutrino oscillations
was found in \cite{Ichikawa:2005vw} to be quite relevant, shifting the lower bound from BBN on the reheating temperature to $2$ MeV.

The radiation content of the Universe can be also tested with observations of 
Cosmic Microwave Background (CMB) anisotropies, the distribution of large-scale structure (LSS) and other
cosmological measurements. This can provide an independent lower bound on $T_{\rm RH}$, as 
discussed in \cite{Kawasaki:1999na}. Using CMB (WMAP) and LSS (2dF) data in combination with BBN, the analysis described
in \cite{Hannestad:2004px} found $T_{\rm RH}> 4$ MeV (95\% CL). A less stringent bound,
$T_{\rm RH}> 2$ MeV (95\% CL), was obtained in \cite{Ichikawa:2006vm} from WMAP data and 
galaxy clustering power spectrum of the SDSS luminous red galaxies, translating the bound on the
radiation content with the relation between $T_{\rm RH}$ and $N_{\rm eff}$. A similar analysis
performed in \cite{DeBernardis:2008zz} found, at 95\% CL, $T_{\rm RH}> 2$ MeV from WMAP-5 data alone
and $T_{\rm RH}> 3.2$ MeV including external priors from cosmic age constraints and 
the SDSS-LRG galaxy survey.

Motivated by the present availability of very precise cosmological data, in this paper we update previous analyses of very low reheating scenarios \cite{Kawasaki:1999na,Kawasaki:2000en,Hannestad:2004px}, and in particular that of ref.\ \cite{Ichikawa:2005vw}, where the effect of flavor neutrino oscillations was included.
We improve previous calculations of the production and thermalization of neutrinos in the low-reheating case 
solving the momentum-dependent equations of motion of the neutrino spectra taking into account three-flavor  oscillations, 
as in our works on the standard case \cite{Mangano:2005cc,new_stand_Neff}
or including neutrino asymmetries \cite{Mangano11,Mangano12}. The impact on BBN is found with a modified version of the
{\sc PArthENoPE} code \cite{Pisanti:2007hk}, while the bounds on $T_{\rm RH}$ from late-time cosmological
observables are obtained from the latest results of the Planck satellite, among other data, including also the case of massive neutrinos.

This paper is organized as follows. We describe our calculations of the 
production of neutrinos in low-reheating scenarios in Sec.\ \ref{sec:evol},
which are used to find the limit on the reheating temperature from
BBN in Sec.\ \ref{sec:BBN}. The bounds on $T_{\rm RH}$
using late cosmological observables, such as CMB anisotropies as measured by Planck,
are described in Sec.\ \ref{sec:CMB}. Finally, in Sec.\ \ref{sec:concl} we
draw our conclusions.

\section{Production of neutrinos in low-reheating scenarios} \label{sec:evol}

Let us call $\phi$ the massive particles that decay with a rate $\Gamma_\phi$ into relativistic particles 
other than neutrinos, reheating the Universe. With this assumption, neutrinos are populated via weak 
interactions with charged leptons such as $e^\pm$. The cases were hadrons or neutrinos can be directly 
produced in $\phi$ decays were considered in 
\cite{Kawasaki:2000en} and \cite{Hannestad:2004px},  respectively. 

We follow the convention of previous analyses and define the reheating temperature $T_{\rm RH}$ as
\begin{equation}
\Gamma_\phi = 3H(T_{\rm RH}) \, ,
\label{Gamma_TRH}
\end{equation}
where we assume that at this point the Universe is already dominated by radiation 
with $T_{\rm RH}$. For the range of temperatures that we are interested in, if all 
relativistic particles are the standard ones with an energy density $\rho_{\rm r}$, 
the Hubble parameter is given by
\begin{equation}
H=\sqrt{\frac{8\pi\rho_{\rm r}}{3 M_P^2}}
=\sqrt{g^*\frac{8\pi^3}{90}}\frac{T^2}{M_P}\, ,
\label{Hrad}
\end{equation}
where $M_P$ is the Planck mass and the number of degrees of freedom would be
$g^*=10.75$ in the standard case. In our low-reheating scenario we use this value 
for the definition of $T_{\rm RH}$, despite the fact that neutrinos might be far from being
in equilibrium with the electromagnetic plasma. Therefore, $T_{\rm RH}$ is just 
a different way of referring to the decay rate of the massive particles
\begin{equation}
T_{\rm RH} \simeq 0.7\left(\frac{\Gamma_\phi}{{\rm s}^{-1}}\right )^{1/2}{\rm MeV}\, .
\label{TRH_Gamma}
\end{equation}

The particle content of the Universe as a function of time in a low-reheating scenario is found 
solving three types of evolution equations: for the $\phi$'s, for the electromagnetic 
particles and for neutrinos. The equation for the energy density of $\phi$'s is that of a 
decaying non-relativistic species
\begin{equation}
\frac{d\rho_\phi}{dt}=-\Gamma_\phi\rho_\phi-3H\rho_\phi \, ,
\label{drhophi}
\end{equation}
where inverse decays are neglected and the Hubble parameter depends on
the total energy density of the Universe.

Those particles that are coupled through electromagnetic interactions are in equilibrium
with a common temperature $T_\gamma$, including photons, $e^\pm$ and $\mu^\pm$.
Therefore, we just need to compute the time evolution of $T_\gamma$, which is obtained from
the continuity equation for the total energy-momentum $\rho$ in the expanding Universe,
\begin{equation}
\frac{d\rho}{dt}=-3H \left (\rho+P\right ) \, ,
\label{rhodot}
\end{equation}
with $P$ the total pressure. The equation for $T_\gamma$ is similar to the one used in
\cite{Kawasaki:1999na,Hannestad:2004px}, i.e.\
\begin{equation}
\frac{dT_\gamma}{dt}=
\frac{-\Gamma_\phi\rho_\phi+4H(\rho_\gamma+\rho_\nu)+3H(\rho_l+P_l)+d\rho_\nu/dt}
{\partial \rho_\gamma/\partial T_\gamma+\partial \rho_{l}/\partial T_\gamma},
\label{dTgammadt}
\end{equation}
where $l$ stands for the charged leptons. In our calculations we have modified this equation
including finite temperature QED corrections to the electromagnetic plasma, as described in
\cite{hep-ph/9702324,astro-ph/0111408}.

The evolution of the momentum distributions of the three flavor neutrinos is calculated taking into
account that, at MeV temperatures, neutrinos are interacting while oscillations start to be effective.
In such a case, the neutrino distributions are described with $3\times3$ matrices in
flavor space $\varrho_{\bf p}$ \cite{Sigl:1993fn, McKellar:1994ja} for each neutrino momentum ${\bf p}$.
The diagonal elements of $\varrho_{\bf p}$ are the usual occupation numbers of flavor neutrinos 
(from which one obtains $\rho_\nu$) and the off-diagonal ones encode phase information, 
vanishing for zero mixing. In the absence of a lepton asymmetry, 
antineutrinos have the same evolution as neutrinos, so there is no separate equation for them.

The equations of motion for $\varrho_{\bf p}$ are the same as in \cite{Pastor:2008ti},
\begin{equation}
\frac{d\varrho_{\bf p}}{dt} ={\rm -i}\,[{\sf\Omega}_{\bf p},\varrho_{\bf
p}]+ C(\varrho_{\bf p})\,.
\label{drhodt}
\end{equation}
where the anticommutator term describes flavor oscillations,
\begin{equation}
{\sf\Omega}_{\bf p}=\frac{{\sf M}^2}{2p}
-\frac{8\sqrt{2}\,G_{\rm F}p}{3 m_{\rm W}^2}\,{\sf E}\,.
\label{vacterm}
\end{equation}
Here $p=|{\bf p}|$ and ${\sf M}$ is the neutrino mass matrix, diagonal in 
the mass basis, whose form in the weak-interaction basis is found using the
neutrino mass matrix $U$ \cite{Mangano:2005cc}. In this work we fix the
neutrino oscillation parameters (two mass-squared differences and three
mixing angles) to the best-fit values found in \cite{Forero:2014bxa} for the
normal mass hierarchy. After the last mixing angle $\theta_{13}$ 
was recently measured with a non-zero value, all of them are known with good precision
(varying them in the allowed ranges does not modify our results).
Matter effects are included via the term
proportional to the Fermi constant $G_{\rm F}$,
where $m_{\rm W}$ is the W boson mass and 
${\sf E}$ is the $3\times3$ flavor matrix of charged-lepton energy densities
\cite{Sigl:1993fn} (we only need to include
the contribution of $e^\pm$ and $\mu^\pm$).

\begin{figure}
\includegraphics[width=0.5\textwidth]{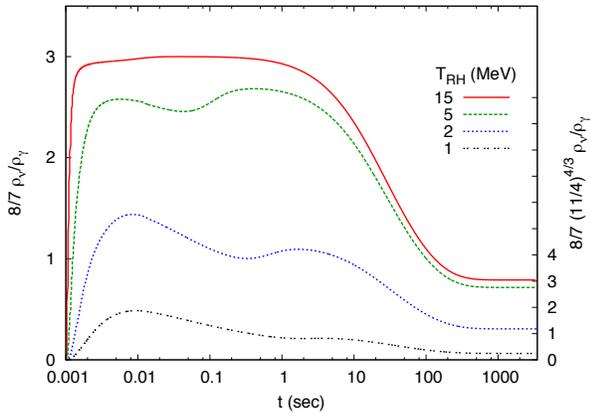}
\caption{Time evolution of the ratio of energy densities of neutrinos and photons, normalized in such a way that it corresponds
to $N_{\rm eff}$ before (left) and after (right) $e^\pm$ annihilations. Four cases with different values of the reheating temperature
are shown.}
\label{fig:evol_rho}
\end{figure}
The last term in eq.\ (\ref{drhodt}) corresponds to the effect of neutrino
collisions, i.e.\ interactions with exchange of momenta. Here we follow
the same assumptions of our previous works in relic neutrino decoupling: we use
momentum-dependent damping factors for the off-diagonal collision terms
in the weak-interaction basis, while scattering and pair processes for the diagonal elements of
$\varrho_{\bf p}$ are included without approximations solving
numerically the collision integrals as in \cite{Mangano:2005cc}. 
For more details on the collision terms and related references, see ref.\ \cite{Pastor:2008ti}.

In summary, with respect to the previous calculation of neutrino production in 
low-reheating scenarios \cite{Ichikawa:2005vw} we include three-flavor oscillations and solve
two-dimensional collision integrals for the weak processes (with $m_e\neq0$
and using Fermi-Dirac distributions for $e^\pm$). For comparison, we have also
performed the calculations in the absence of neutrino oscillations. As in
\cite{Ichikawa:2005vw} we neglect neutrino-neutrino processes, that do not
increase the energy density of neutrinos and are expected to have small effects,
and use comoving variables for the expansion rate ($x=m_ea$), the photon temperature
($z=T_\gamma a$) and the neutrino momenta ($y=p_\nu a$), where $a$ is the scale factor.

For each value of the reheating temperature, we start our calculations at $t=10^{-3}$ s, 
imposing that the Universe is strongly dominated by matter, and end the evolution
when $T_\gamma <10$ keV in a radiation-dominated regime. The results do not depend on the
choice of the initial time, provided that the maximum value of $T_\gamma$ that is reached
is significantly larger than the neutrino decoupling temperature. The value of $\rho_\phi$ is obtained 
from the Friedmann equation, $3t_i/2=(8\pi\rho_\phi/3M_P^2)^{-1/2}$ \cite{Hannestad:2004px}, while
we fix $T^0_\gamma=2$ MeV. This last choice speeds up the initial phase of the numerical calculations,
until the expected evolution is obtained: $T_\gamma$ decreases as $t^{-1/4}$ when matter dominates
and as $t^{-1/2}$ when relativistic particles fix the cosmological expansion (see figure 1 in \cite{Kawasaki:2000en}).

\begin{figure}
\includegraphics[width=0.5\textwidth]{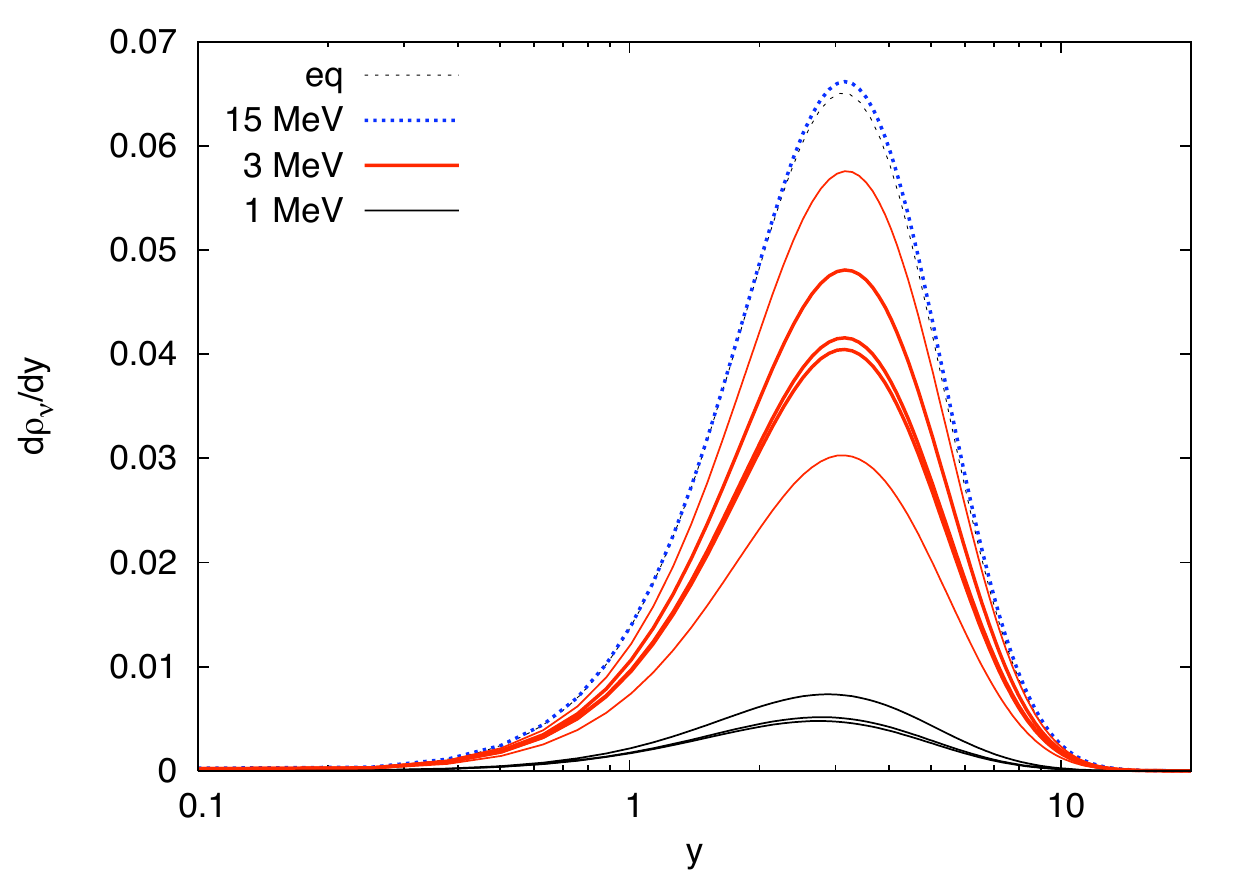}
\caption{Final differential spectra of neutrino energies as a function of the comoving momentum for 
three values of the reheating temperature, compared to an equilibrium spectrum (thin dotted black line). 
The three thick solid lines for $T_{\rm RH}=3$ (middle red lines) and $1$ MeV (lower black lines) correspond, from
larger to smaller values, to $\nu_e$, $\nu_\mu$ and $\nu_\tau$, respectively. For $T_{\rm RH}=3$ MeV we also
include the case without flavor oscillations (thin red lines, upper for $\nu_e$ and lower for $\nu_{\mu,\tau}$).}
\label{fig:drho}
\end{figure}
A few examples of our numerical calculations are depicted in Fig.\ \ref{fig:evol_rho}, where the time evolution
of the ratio of energy densities $\rho_\nu/\rho_\gamma$ is shown for four values of $T_{\rm RH}$. This ratio is normalized in two 
different ways in the y-axis, so that it corresponds to the effective number of neutrinos before and after $e^\pm$ annihilations 
(3 in both cases in the approximation of instantaneous neutrino decoupling) \cite{Pastor:2008ti}. In
particular, at late times $N_{\rm eff}$ is defined as $\rho_{\rm r}/\rho_\gamma=1 + 7/8 (4/11)^{4/3} N_{\rm eff}$. For 
values of the reheating temperature as large as $15$ MeV, except at the very initial phase, the evolution of $\rho_\nu/\rho_\gamma$ 
is similar to the standard case with a fast drop at $t\sim 1$ s due to photon heating by $e^+e^-$ annihilations. Instead,
for smaller values of $T_{\rm RH}$ one can see that $\rho_\nu/\rho_\gamma$ decreases while the $\phi$'s decay, followed
by a period where this ratio slightly increases but never reaches the value expected in the standard case.  
Instead, neutrinos do not reach equilibrium and the final $N_{\rm eff}$ is clearly below three.
This fact can be easily seen in Fig.\ \ref{fig:drho}, where the differential energy spectrum of neutrinos at the end of the evolution 
is shown for three values of $T_{\rm RH}$, compared to the case of an equilibrium Fermi-Dirac spectrum. For $T_{\rm RH} =15$ MeV
the energy distribution is very close to equilibrium, actually slightly above for large momenta as expected in the
standard decoupling case that leads to $N_{\rm eff}=3.046$ \cite{Mangano:2005cc}. Instead, for smaller values
of the reheating temperature there exists a significant reduction in the production of neutrinos, in particular for 
$T_{\rm RH} =1$ MeV.

\begin{figure}
\includegraphics[width=0.42\textwidth]{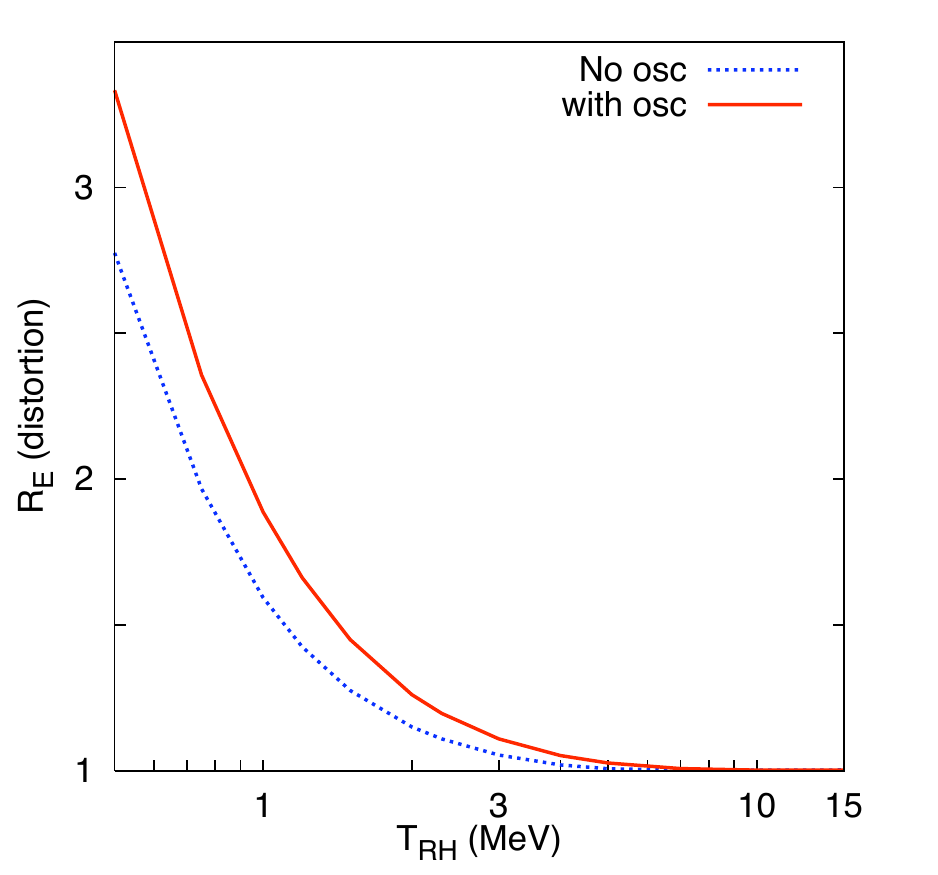}
\caption{Distortion of the electron neutrino spectrum parameterized with $R_E$ (defined in the text) 
as a function of the reheating temperature. A value $R_E>1$ indicates a significant spectral distortion with
respect to equilibrium. Solid curve is for oscillating neutrinos, while dotted is for the no oscillation case and is reported for comparison.}
\label{fig:R_E}
\end{figure}
The distortion of the neutrino spectra from an equilibrium form can be described by the parameter
$R_E$, defined as \cite{Kawasaki:2000en}
\begin{equation}
R_E=\frac{1}{3.151\,T_{{\rm eff},\nu}}\frac{\rho_\nu}{n_\nu} \, ,
\label{R_E}
\end{equation}
where the neutrino energy density $\rho_\nu$, number density $n_\nu$ and effective
temperature $T_{{\rm eff},\nu}=[2\pi^2n_\nu/\zeta(3)]^{1/3}$
are found by integrating the spectrum. The parameter $R_E$ represents the ratio of 
the mean neutrino energy to the value in thermal equilibrium and it is shown for
electron neutrinos as a function of $T_{\rm RH}$ in Fig. \ref{fig:R_E}. A value around unity 
indicates that the neutrino spectrum has a form close to equilibrium, while larger
$R_E$ point to significant distortions as happens for  $T_{\rm RH}\lesssim 3$ MeV (more
noticeable if flavor oscillations are included).

\begin{figure}
\includegraphics[width=0.45\textwidth]{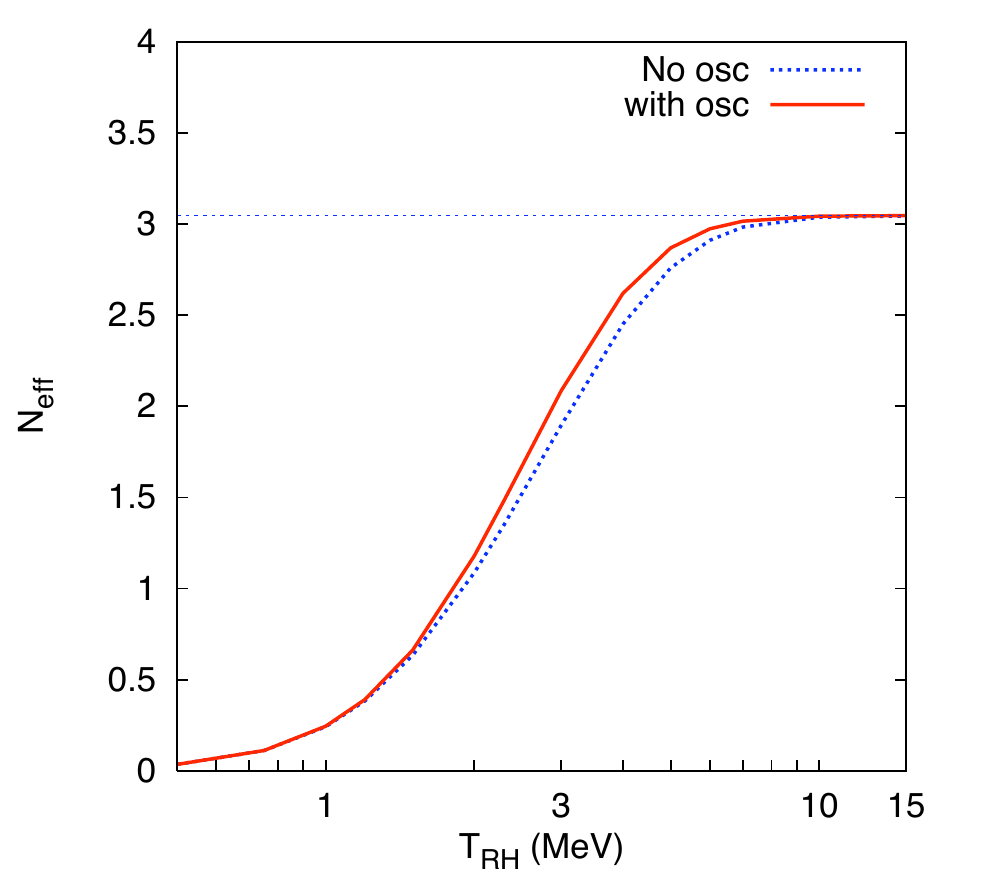}
\caption{Final contribution of neutrinos to the radiation energy density in terms of $N_{\rm eff}$, as a function of the reheating temperature. The
horizontal line indicates the standard value, $N_{\rm eff}=3.046$.}
\label{fig:neffTRH}
\end{figure}

Finally, we present in Fig.\  \ref{fig:neffTRH} the final contribution of neutrinos to the energy density of 
radiation in terms of $N_{\rm eff}$. Its value drops below $3$ if $T_{\rm RH}\lesssim 7$ MeV.
Our results are similar to those shown in \cite{Hannestad:2004px} (figure 1) and \cite{Ichikawa:2005vw} (figure 3). 
In particular, we agree with ref.\ \cite{Ichikawa:2005vw} in the fact that
the final value of $N_{\rm eff}$ in the middle range of MeV reheating temperatures is slightly larger in the case where flavor neutrino oscillations are included.

\section{Bounds from Primordial Nucleosynthesis} \label{sec:BBN}

The effect of neutrino distributions on the production of primordial yields of light elements during BBN can be summarized as follows:
\begin{enumerate}
\item
the energy density of all neutrino flavors $\rho_\nu$, contributes to the radiation energy density $\rho_{\rm r}$, which leads the Hubble expansion rate;
\item
the momentum distribution of electron neutrinos directly enters the charged current weak rates, which govern the neutron-proton chemical equilibrium;
\item
the time evolution of $\rho_\nu$ explicitly enters eq.\ (\ref{rhodot}), stating the entropy conservation per comoving volume; in particular, one can define the function \cite{Serpico:2004gx}
\begin{equation}
\mathcal{N}=\frac{1}{z^4}\left(x \frac{d}{dx}\bar\rho_\nu\right)\, ,
\label{naux}
\end{equation}
with $\bar\rho_\nu\equiv a^4 \rho_\nu$. This function measures neutrino heating by electromagnetic plasma, and in the standard scenario is not vanishing during the $e^\pm$ annihilation stage only, due to the small entropy transfer (order percent) to neutrinos. For sufficiently low $T_{\rm RH}$ scenarios we expect this function to be non zero in the BBN relevant temperature range, during the neutrino production stage.
\end{enumerate}
\begin{figure}[b]
\includegraphics[height=0.26\textwidth]{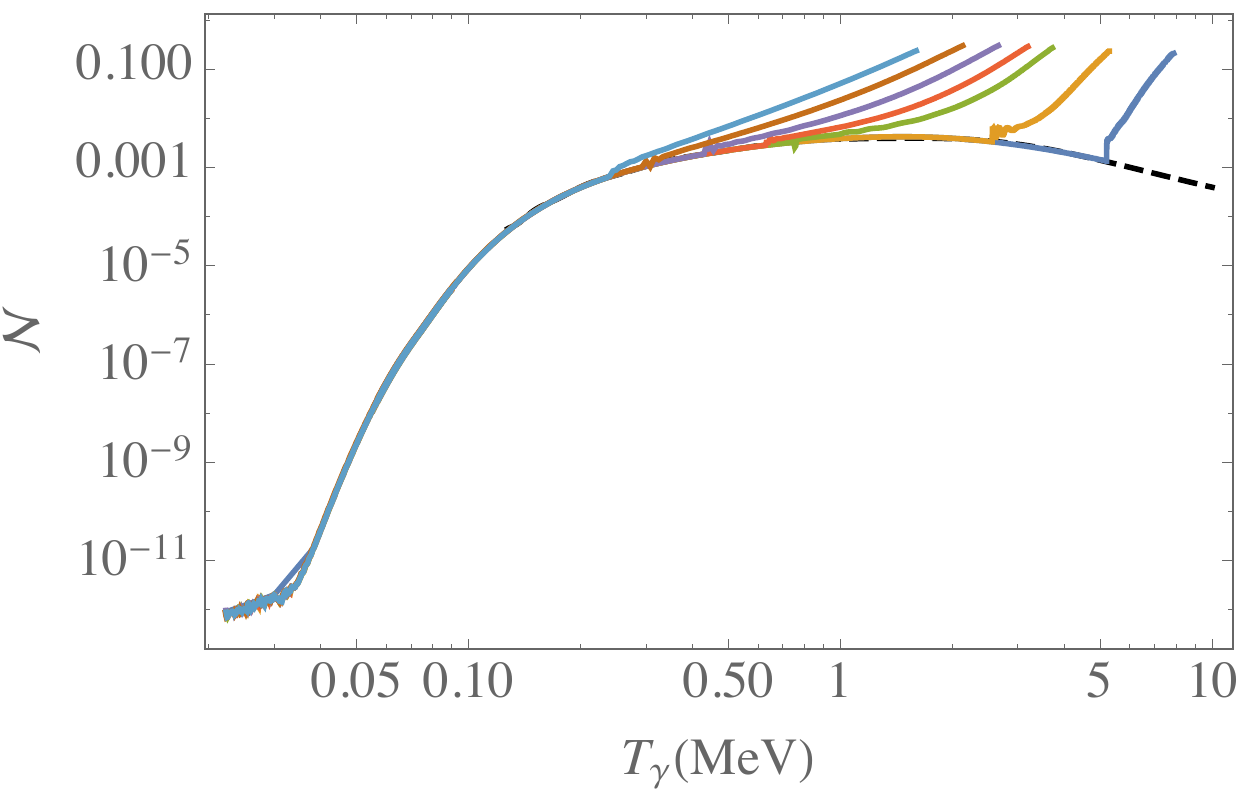}
\caption{The function $\mathcal{N}$ in eq.\ (\ref{naux}) for different values of $T_{\rm RH}$ (15, 10, 7, 6, 5, 4, 3 MeV from right to left) compared to the standard one (black dashed line). The initial (large temperature) decreasing branch of each curve is due to neutrino heating from electromagnetic plasma, which smoothly tends toward the standard behavior due to $e^\pm$ annihilation stage.}
\label{fig:naux}
\end{figure}

In order to put constraints on the reheating temperature, the time evolution of $\rho_\nu$, $\mathcal{N}$, and of the $\nu_e$ ($\bar\nu_e$) distribution functions have been fitted for different values of $T_{\rm RH}$, and used as input in a modified version of the {\sc PArthENoPE} code \cite{Pisanti:2007hk}. Fig.\  \ref{fig:naux} shows the evolution with photon temperature of  $\mathcal{N}$ for different values of $T_{\rm RH}$. As expected, the lower is $T_{\rm RH}$ the smaller the value of $T_\gamma$ at which the standard expression is recovered.

\begin{figure}[t]
\includegraphics[width=0.45\textwidth]{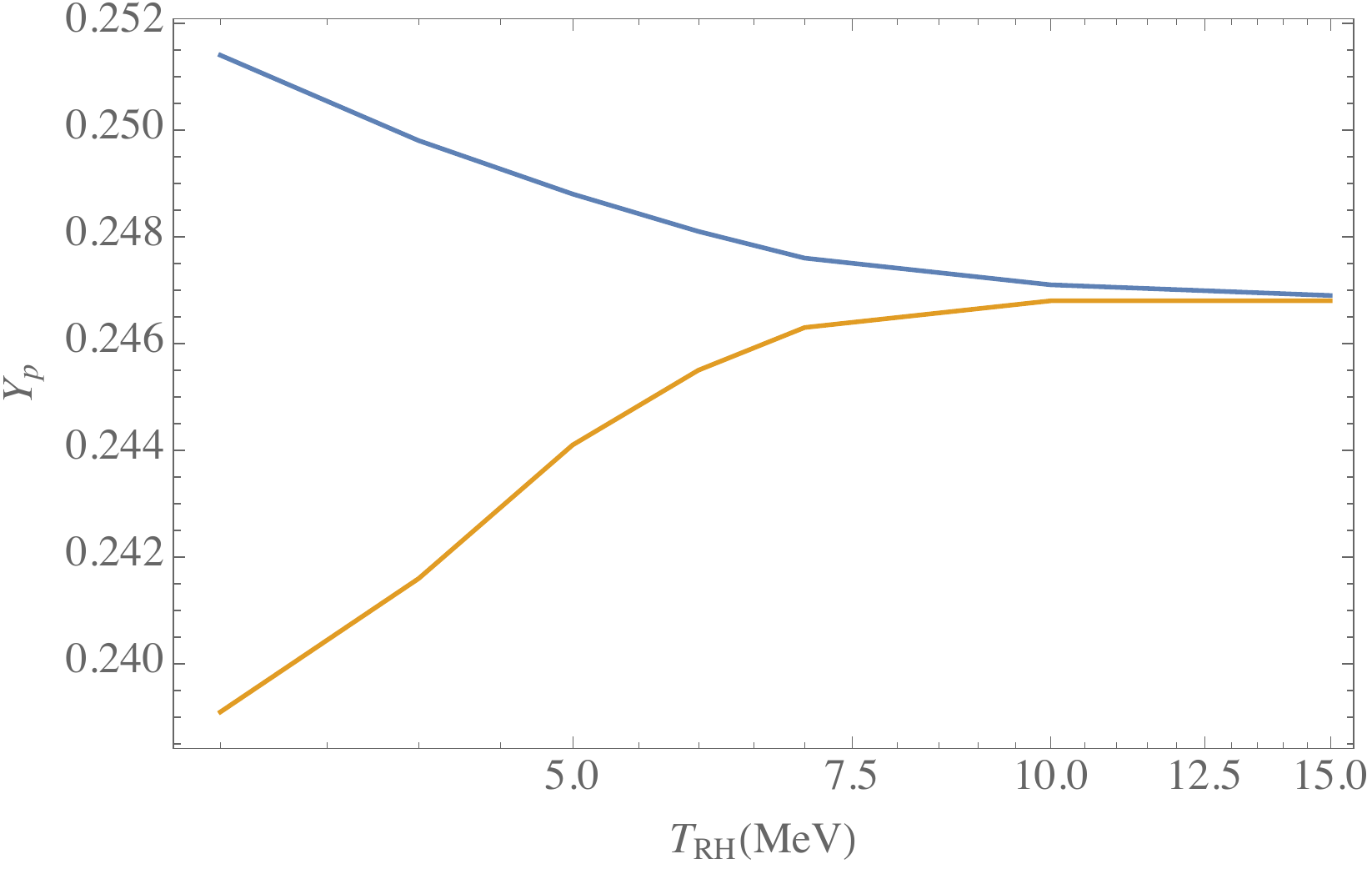}
\caption{Values of the primordial helium yield, $Y_p$, for different values of $T_{\rm RH}$, taking into account neutrino oscillations (upper blue line) and in absence of the oscillations (lower yellow line).}
\label{fig:he4}
\end{figure}

\begin{figure}[b]
\includegraphics[width=0.45\textwidth]{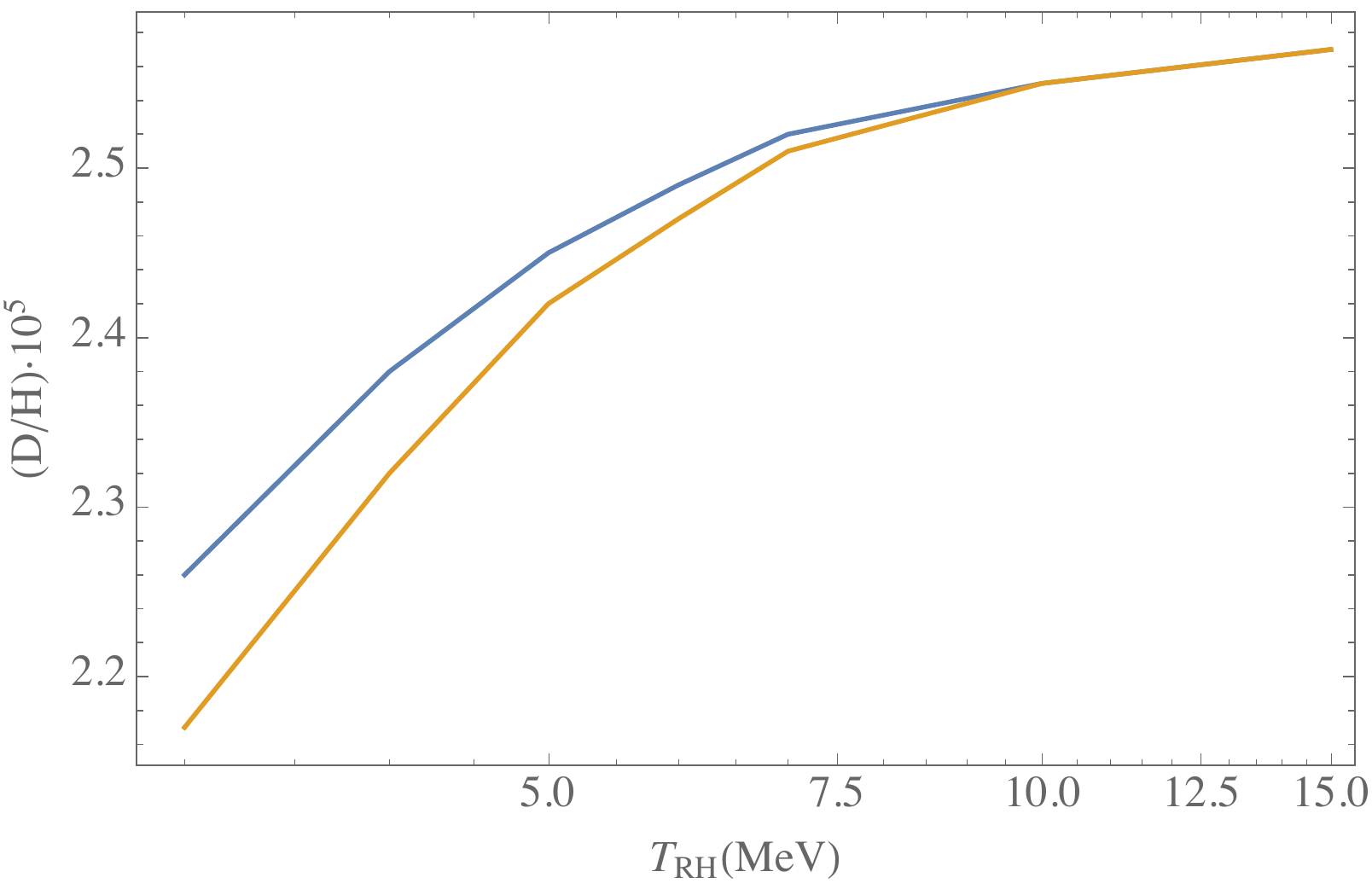}
\caption{Values of the deuterium to hydrogen ratio D/H, as a function of $T_{\rm RH}$, with and without neutrino oscillations (upper blue and lower yellow lines, respectively).}
\label{fig:h2}
\end{figure}

The values of the final abundances of helium, $Y_p$, and deuterium, $^2$H/H, for a baryon density $\omega_b=\Omega_b h^2 = 0.02226$ are reported in Fig.s\  \ref{fig:he4} and \ref{fig:h2}, where we show for comparison the data obtained both in presence and in the absence of neutrino oscillations. In particular, the dependence on $T_{\rm RH}$ of the helium abundance of Fig.\  \ref{fig:he4} is similar to what was found in \cite{Ichikawa:2005vw} (figure 4).


We use the most recent data on the primordial abundances of $^4$He and deuterium. For the helium mass fraction, the result of new observations of He (and H) emission lines in extragalactic HII regions, including a near infrared line at $\lambda$10830 \cite{Izotov:2014fga} that helps in breaking the degeneracy between gas density and temperature, leads to a reduction in the uncertainty and to a better defined regression value \cite{Aver:2015iza},
\begin{equation}
Y_p=0.2449\pm0.0040\,\,\, .
\label{yp}
\end{equation}
On the other hand, after a very precise observation of deuterium abundance in 2012 \cite{Pettini:2012ph}, which reduced the uncertainty from $10-20$\% to about 2\%, the result of a reanalysis of deuterium data gives \cite{Cooke:2013cba}
\begin{equation}
^2\mbox{H/H}=(2.53\pm 0.04)\cdot 10^{-5}\,\,\, .
\label{h2}
\end{equation}
For each nuclide we have defined a $\chi^2$-function as follows ($i=$ $^4$He , $^2$H/H)
\begin{equation} 
\chi^2_i = \frac{(X_i^{th}(\Omega_b h^2, T_{\rm RH})-X_i^{exp})^2}{\sigma^2_{i,exp}+\sigma^2_{i,th}}
\end{equation}
where $X_i^{th}$ is the theoretical value computed with {\sc PArthENoPE} code as function of baryon density and $T_{\rm RH}$, and $\sigma^2_{th}$ the corresponding uncertainty due to propagation of nuclear process rates ($\sigma_{^4\mbox{\scriptsize He},th}= 0.0003$, $\sigma_{\mbox{\scriptsize D},th}=0.07$). The corresponding experimental mean value and uncertainty are denoted by $X_i^{exp}$ and $\sigma_{i,exp}$. We have used a prior on the value of $\Omega_b h^2$ from the recent Planck collaboration results  \cite{Ade:2015xua}. Of course each $\chi^2_i$ tends to a constant value for sufficiently large values of $T_{\rm RH}$, which corresponds to a standard cosmology at BBN epoch, whereas it increases quite rapidly when $T_{\rm RH}$ becomes smaller and smaller. Marginalizing the product $\left(\chi^2_{^4\mbox{\scriptsize He}}\chi^2_{\mbox{\scriptsize D}}\right)(\Omega_b h^2, T_{\rm RH})$ over the parameter $\Omega_b h^2$, we find that the ``$2 \sigma$" BBN bound on the reheating temperature, obtained by requiring that the marginalized $\left(\chi^2_{^4\mbox{\scriptsize He}}\chi^2_{\mbox{\scriptsize D}}\right)(T_{\rm RH}) \leq 4$, is
\begin{equation}
T_{\rm RH} \geq 4.1 \, \mbox{MeV} \,\,\, .
\end{equation}
Actually, this constraint is entirely provided by deuterium, whose experimental estimate is more accurate. 

As a final consideration, we mention that our result was obtained using the present best fit of experimental data on the $d(p,\gamma)^3$He S-factor 
\cite{Adelberger:2010qa}. A theoretical ab-initio calculation for this process has been presented in \cite{Marcucci:2005zc}, resulting in a larger reaction cross section in the BBN energy range, and a lower theoretical value of D/H, in better agreement with the experimental result of eq.\ (\ref{h2}), see \cite{DiValentino:2014cta}. Using this theoretical value for the $d(p,\gamma)^3$He thermal rate, assuming conservatively the same error on its determination $\sigma^2_{\mbox{\scriptsize D},th}=0.07$, the BBN bound on the reheating temperature becomes even stronger
\begin{equation}
T_{\rm RH} \geq 5.1 \, \mbox{MeV} \,\,\, .
\end{equation}

\section{Bounds from Cosmic Microwave Background observations} \label{sec:CMB}

We now derive bounds on the reheating temperature from the observed spectrum of CMB temperature and polarization
anisotropies as measured by the Planck satellite \cite{Adam:2015rua,Aghanim:2015xee,Ade:2015xua}.

In order to compute the CMB power spectrum
in models with a low-reheating temperature, we have modified the Boltzmann code \verb+CAMB+ \cite{Lewis:1999bs}
to allow for an arbitrary form of the neutrino distribution function (in principle different for each neutrino state).
We use the results on the evolution of the neutrino spectra, as described in Sec. II above, to obtain the distribution 
functions of the flavor neutrinos, to be given as an input to \verb+CAMB+ for a given
value of the reheating temperature $\Trh$. Since at the redshift of interest for the calculation of
the CMB anisotropies and other late-time observables the neutrino distribution
functions are evolving self-similarly, keeping a constant shape and being only redshifted by the expansion of the Universe,
only the very last step of the time integration described in Sec. II is actually relevant to this purpose. 
Note, however that the cosmological perturbation equations are written in terms of the momentum distributions $f_{\nu_i}\,(i=1,\,2,\,3)$ for the mass eigenstates;
these are related to the flavor distributions $f_{\nu_\alpha}\,(\alpha=e,\,\mu,\,\tau)$ by means of the relation:
\be
f_{\nu_i} (y) = \sum_{\alpha = e,\,\mu,\,\tau} \left|U_{\alpha i}\right|^2 f_{\nu_\alpha}(y)
\ee
where $U$ is the neutrino mixing matrix. As already said, we fix the matrix elements to the best-fit values from the global analysis of oscillation data of Ref. \cite{Forero:2014bxa} for the normal mass hierarchy.

We have seen from the results of Sec. \ref{sec:evol} that in low-reheating scenarios the effective number of relativistic species $\Neff$ 
is smaller than its standard value. This is not an independent parameter as it only depends on the neutrino distribution functions and thus
on the reheating temperature. So, the correct value of $\Neff$ for each $\Trh$ is automatically obtained internally in the Boltzmann code
by means of the integration of the neutrino spectra. 

The CMB anisotropy spectrum is also sensitive to the primordial helium abundance $Y_p$,
mainly through its influence on the recombination history. In particular, increasing the helium
fraction has the effect of reducing the power in the damping tail of the CMB spectrum. Assuming standard BBN,
the helium fraction is not an independent parameter but instead can be determined
once the baryon density $\omega_b$ and the effective number of relativistic species $N_\mathrm{eff}$ 
- related to the expansion speed at the time of BBN - 
are given. Thus, in current cosmological analyses that assume standard BBN, the abundance of helium 
is consistently computed from $\omega_b$ and $N_\mathrm{eff}$. 
We generalize this treatment to the models with a low-reheating temperature, by using \verb+PArthENoPE+
to generate a grid of values of $Y_p$ as a function of $\omega_b$ and $\Trh$, from which we interpolate to obtain the 
helium abundance for arbitrary values of these two parameters.

Finally, in order to compute Bayesian confidence intervals and sample the posterior distributions
for the parameters of the model, given the data, we use
the Metropolis-Hastings (MH) algorithm as implemented in
\verb+CosmoMC+ \cite{Lewis:2002ah} (interfaced with our modified version of
\verb+CAMB+). The MH algorithm is used to generate Markov chains of samples for a 
set of cosmological parameters. The models under investigation here can be described by the values of the six parameters
of the standard $\Lambda$CDM model, namely
the present density parameters
$\Omega_\mathrm{b} h^2$ and $\Omega_\mathrm{dm} h^2$ of baryons and
dark matter respectively, the angular size of the sound horizon at recombination
 $\theta$, the optical depth
to recombination $\tau_\mathrm{rec}$, the spectral index $n_s$ and
amplitude $A_s$ (evaluated at the pivot scale $k_0=0.05$ Mpc$^{-1}$)
of the spectrum of primordial scalar fluctuations, with the addition
of the reheating temperature $\Trh$ and of the mass $m_1$ of the lightest 
neutrino eigenstate. The masses of the remaining eigenstates are derived
assuming a normal hierarchy (for the mass differences we again use the results of Ref. \cite{Forero:2014bxa}), 
and we assume spatial flatness and 
adiabatic initial conditions. We take flat priors on all the parameters of the model; in particular, we
take $3\,\MeV \le \Trh \le 15\,\MeV$.
As explained above, for each set of parameter values - corresponding to a step in the Monte Carlo - 
we use interpolation tables to obtain the values of the helium abundance $Y_p$ 
and of the parameters of the neutrino distribution functions to be fed to \verb+CAMB+
along with the other parameters. 

We compute parameter constraints from different datasets. Our basic dataset consists of Planck 2015 data
on the CMB temperature anisotropies up to a multipole $\ell = 2500$ and on large-scale ($\ell < 30$) polarization anisotropies
\cite{Aghanim:2015xee}. 
This is the same basic dataset used
by the Planck collaboration for parameter estimation \cite{Ade:2015xua}, and we will follow
the custom to refer to this dataset as ``PlanckTT+lowP''. We will also consider  the ``PlanckTTTEEE+lowP''  dataset that includes, 
in addition to the data just mentioned, the high-ell polarization measurements from Planck \cite{Aghanim:2015xee}.

The likelihood functions associated to the datasets just described, are evaluated 
and combined using the likelihood code
distributed by the Planck collaboration \cite{Aghanim:2015xee}, and publicly
available at Planck Legacy Archive\footnote{\url{http://pla.esac.esa.int/pla/}}.
A number of additional ``nuisance'' parameters, required to describe 
e.g., unresolved foreground components and instrumental characteristics,
are marginalized over.

As a first step, we have performed a test run by fixing $\Trh$ to 15 MeV. For this value of the reheating 
temperature, the scenario should be basically indistinguishable from standard $\Lambda$CDM, and in fact
we have checked that we can reliably reproduce the corresponding results presented in Ref. \cite{Ade:2015xua}.
We indeed find small (below one $\sigma$) shifts in the parameters, but we have traced them to the different 
parameterization of neutrino masses between our code and the standard version of \texttt{CosmoMC} used
by the Planck collaboration (see below).

\begin{figure}[tbp]
\begin{center}
\includegraphics[width=0.95\linewidth,keepaspectratio]{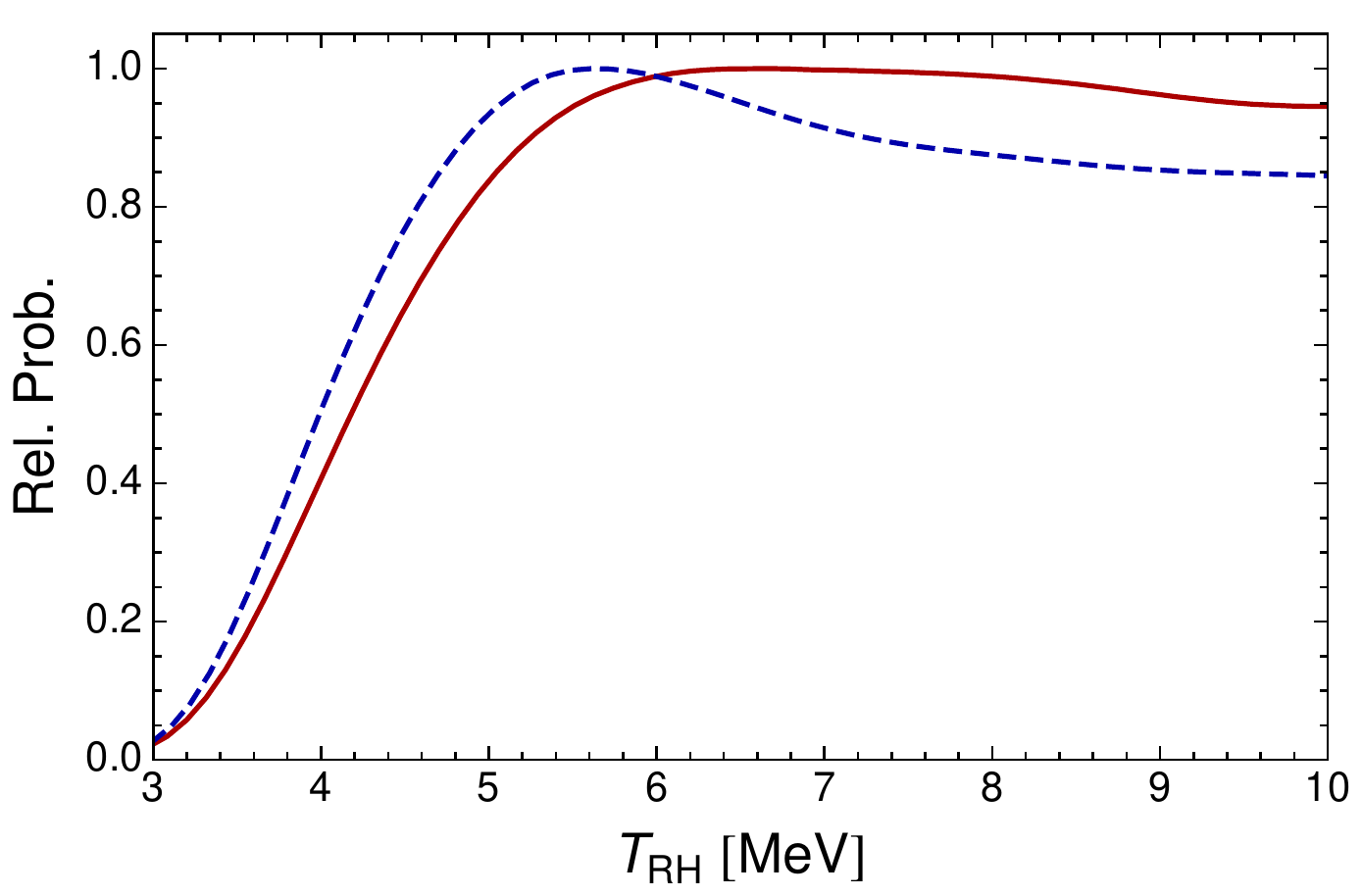}
\caption{One-dimensional probability distribution for $\Trh$, from the PlanckTT+lowP (red solid) and PlanckTTTEEE+lowP (blue dashed) datasets.}
\label{fig:post1D_TRH}
\end{center}
\end{figure}

Then we turn to the Monte Carlo runs with $\Trh$ as a free parameter.
We get the following $95\%$ lower limit on the reheating temperature
\begin{align}
&\Trh \ge 4.7\,\MeV \, \quad (\mathrm{PlanckTT+lowP}) \,,\nonumber \\
&\Trh \ge 4.4\,\MeV \, \quad (\mathrm{PlanckTTTEEE+lowP}) \,.
\end{align}
The corresponding posterior distributions for $\Trh$ are shown in Fig. \ref{fig:post1D_TRH}.
For what concerns the effective number of relativistic
degrees of freedom, we get (again at $95\%$ C.L.)
\begin{align}
&\Neff \ge 2.81  \quad (\mathrm{PlanckTT+lowP}) \, , \nonumber \\
&\Neff \ge 2.75  \quad (\mathrm{PlanckTTTEEE+lowP}) \, .
\end{align}
These results are in agreement with the fact, reported in the Planck parameters paper \cite{Ade:2015xua},
that the addition of high-ell polarization data tends to slightly shift $N_\mathrm{eff}$ towards lower values.

It is interesting to test the robustness of the cosmological limits on neutrino masses in low-reheating scenarios.
In the following we will focus on the PlanckTT+lowP dataset only.
Computing the $95\%$ credible interval for the  sum of neutrino masses $\sum m_i$ from our chains
yields
\be
\sum m_i \le 0.83\,\eV \qquad(\textrm{for } \Trh \le 15\,\MeV) \,.
\ee
We have checked that we basically obtain\footnote{
This limit is different from the one quoted in the Planck parameters paper  \cite{Ade:2015xua}
for the $\Lambda$CDM+$m_\nu$ scenario ($\sum m_i \le 0.72\,\eV$). This is because we have implemented the
normal mass hierarchy in the code, and as a consequence the total mass is bound to be larger than $0.06\,\eV$. 
Instead, the $\Lambda$CDM+$m_\nu$ analysis 
by the Planck collaboration assumes three degenerate mass states and a simple positivity prior $\sum m_i \ge 0$.}
the same result ($\sum m_i \le 0.80\,\eV$)
when $\Trh$ is fixed to 15 MeV.
However, one should not conclude from this that the neutrino mass limits stay unchanged in a low-reheating scenario.
The reason is that, by considering the full allowed range for $\Trh$, we are \emph{de facto} exploring
a parameter space that for the most part (let us say, for $\Trh > 7\,\MeV$) reproduces the standard
$\Lambda$CDM+$m_\nu$ scenario. This is also the region where most of the probability mass is concentrated. 
When marginalizing over all other parameters to obtain the posterior for $m_\nu$, this region dominates
the probability integral and thus the procedure returns a constraint that is very close to the one found in standard $\Lambda$CDM.
A possibly more sensible way to assess the effect of low-reheating scenarios on cosmological neutrino mass limits is to 
focus on the models with the lowest reheating temperature, by assuming a more stringent prior on $\Trh$ (\emph{e.g.},  $\Trh \le 7\,\MeV$, or $\Trh = 6\,\MeV$).
Assuming different \emph{a priori} upper limits on $\Trh$, we get the following $95\%$ credible intervals for $\sum m_i$:

\be
\sum m_i \le \left\{ 
\begin{array}{cl}
0.89\, \eV &  \quad (\Trh \le 7\,\MeV) \\[0.1cm]
0.93\, \eV &  \quad (\Trh \le 6\,\MeV) \\[0.1cm]
0.96\, \eV &  \quad (\Trh \le 5\,\MeV)
\end{array}
\right.
\ee

\begin{figure}[tbp]
\begin{center}
\includegraphics[width=0.95\linewidth,keepaspectratio]{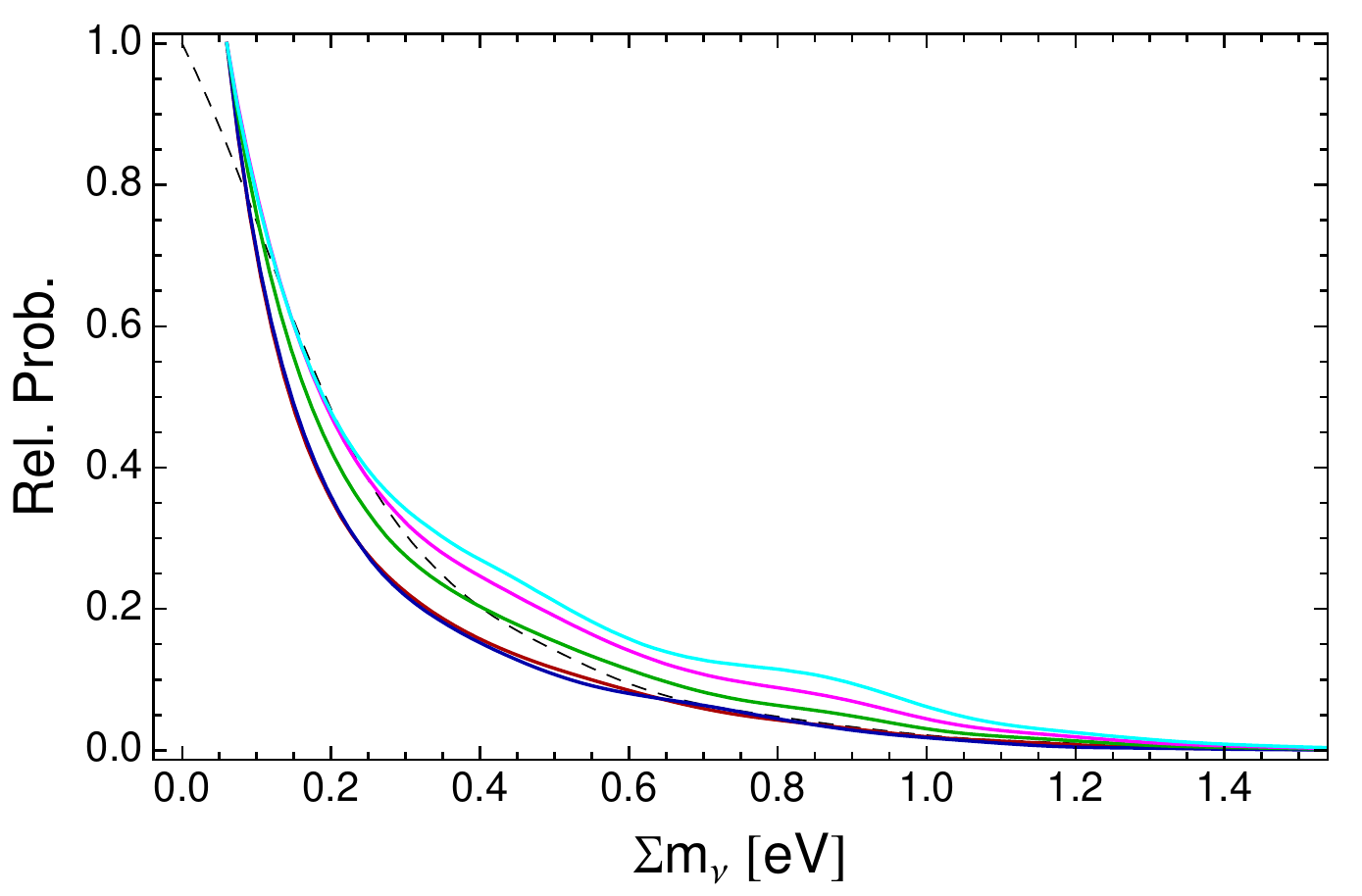}
\caption{One-dimensional posterior distribution for $\sum m_i$ from the PlanckTT+lowP dataset, for different assumptions on $\Trh$:
$\Trh \le 15 \,\MeV$ (red), $\Trh = 15 \,\MeV$ (blue, basically coinciding with the previous case), $\Trh \le 7$ MeV (green),
$\Trh \le 6$ MeV (magenta), $\Trh \le 5$ MeV (cyan). We also show a curve (black dashed) that reproduces
the Planck collaboration results on $\sum m_i$ for the $\Lambda$CDM+$m_\nu$ model, for three 
neutrinos with equal mass.}
\label{fig:post1D_mnu}
\end{center}
\end{figure}

In Fig. \ref{fig:post1D_mnu} we show the posterior distribution for the sum of neutrino masses with these and other
assumptions on $\Trh$.
There is a clear trend here: the constraint relaxes for lower reheating temperatures. The reason is easy to understand: the lower the reheating temperature,
the lower the neutrino number density. Since, within a good approximation, the CMB directly constraints the
 neutrino energy density, and thus the product mass times number density,
a low reheating temperature allows larger masses to be compatible with the data. 

Finally, we have calculated the improvement in the goodness-of-fit of the low-reheating scenario with respect to standard $\Lambda$CDM. For the PlanckTT+lowP dataset, we find $\Delta\chi^2 \ll 1$, 
signaling that low-reheating scenarios, in spite of the presence of one additional parameter, do not allow to improve significantly the fit to the CMB data.

\section{Conclusions} \label{sec:concl}

In this paper we have studied the production of relic
neutrinos in a generic cosmological model where the
latest reheating phase of the Universe occurs at temperatures as low as $1$ MeV.
This low-reheating scenario is an exotic possibility where it is not easy to account for 
a proper baryogenesis (although some solutions exist \cite{Benakli:1998ur,Kohri:2009ka}), but it provides 
an interesting way to reduce the radiation content of the Universe ($N_{\rm eff}$), leaving
room for relativistic particles whose abundance is quite constrained in the standard case, 
such as sterile neutrinos \cite{Gelmini:2004ah,Yaguna:2007wi,Gelmini:2008fq}. The cosmological production of other possible
particles is also modified for very low reheating temperatures, for instance in the case of axions \cite{Grin:2007yg,Visinelli:2009kt}.

We have carefully solved the thermalization of neutrinos in a low-reheating scenario, improving previous calculations
and taking into account the effect of three-flavor neutrino oscillations. We have calculated
the impact on the production of light elements, which in the
case of $^4$He strongly depends on the inclusion or not of neutrino oscillations, 
as originally found in \cite{Ichikawa:2005vw}. The BBN lower limit on the reheating temperature is
$T_{\rm RH} {\rm (BBN)} \geq 4.1$ MeV (95\% CL), and it is fixed by the observed abundance of primordial deuterium. This value is
larger than the BBN bounds found in previous analyses \cite{Kawasaki:1999na,Kawasaki:2000en,Giudice:2000ex,Giudice:2000dp,Hannestad:2004px}.

A slightly more stringent bound on the reheating temperature can be obtained from the analysis of CMB anisotropies.
With the same basic dataset used by the Planck collaboration for parameter estimation, we find the lower limit
$T_{\rm RH} {\rm (CMB)} \geq 4.7$ MeV (95\% CL). At the same time, the cosmological constraints on the total 
neutrino mass are quite robust in a low-reheating scenario, although, as expected, the bounds from Planck are slightly relaxed for
values of $T_{\rm RH}$ below $10$ MeV.

\section*{Acknowledgments}
ML acknowledges support from ASI through ASI/INAF Agreement I/072/09/0 for the Planck LFI Activity of Phase E2. 
Part of this work was carried out while ML was visiting the Instituto de F\'isica Corpuscular in Valencia, whose hospitality is kindly acknowledged, supported by the grant \textit{Giovani ricercatori} of the University of Ferrara, financed through the funds Fondi 5x1000 Anno 2010 and Fondi Unicredit 2013.
GM, GM and OP acknowledge support by
the {\it Instituto Nazionale di Fisica Nucleare} I.S. TASP and the PRIN
2012 ``Theoretical Astroparticle Physics'' of the
Italian {\it Ministero dell'Istruzione, Universit\`a e Ricerca}. 
SP and PFdS were supported by the Spanish grants 
FPU13/03729, FPA2014-58183-P, Multidark CSD2009-00064 and SEV-2014-0398
(MINECO), and PROMETEOII/2014/084 (Generalitat Valenciana). This research was also supported by a
Spanish-Italian MINECO-INFN agreement, ref.\ AIC-D-2011-0689.


\end{document}